\documentclass[12pt]{article}

 \usepackage{amsfonts}   \usepackage{amssymb,amsmath}
 \usepackage{graphicx} \usepackage{epstopdf}
\newcommand{\be}{\begin{equation}}
\newcommand{\ee}{\end{equation}}
\newcommand{\ba}{\begin{eqnarray}}
\newcommand{\ea}{\end{eqnarray}}
\newcommand{\bi}{\begin{itemize}}
\newcommand{\ei}{\end{itemize}}
\newcommand{\la}{\label}
\newcommand{\non}{\nonumber}
 
 \newcommand{\crlb}[1]{\label{#1}\\[2pt]}
 
 \newcommand{\crld}[1]{\label{#1}}
 \newcommand{\eela}[1]{\quad\hbox{\scriptsize{#1}}\label{#1}\end{eqnarray}}
 \newcommand{\eelb}[1]{\label{#1}\end{eqnarray}}
 
 \newcommand{\newsecb}[2]{\section{#1}\label{#2}\setcounter{equation}{0}}

 \newcommand{\nolabels} {\def\eel{\eelb} \def\crl{\crlb} \def\newsecl{\newsecb}\def\bibiteml{\bibitem}\def\citel{\cite}\def\labell{\crld}}

\newcommand\publishversion{\nolabels\setlength{\textheight}{8.75in}\setlength{\oddsidemargin}{0in}
    \setlength{\textwidth}{6.3in}\setlength{\topmargin}{-0.3in}}

 \def\Tr{{\mbox{Tr}}\,}                
        \def\be{\begin{eqnarray}}    \def\ee{\end{eqnarray}}
 \def\bi#1{\begin{itemize}\item[#1]}     \def\ei{\end{itemize}}  
   \def\^#1{\hat{#1}}

       \def\b{\beta}         
 \def\d{\delta}      \def\D{\Delta}   
       \def\l{\lambda} \def\L{\Lambda}     \def\m{\mu}
                 
                  \def\s{\sigma}

     \def\OO{{\mathcal O}}

 \def\ffract#1#2{\raise .2 em\hbox{$\scriptstyle#1$}\kern-.3em/
                 \kern-.2em\lower .15 em \hbox{$\scriptstyle#2$}}

\def\bmatrix{\begin{matrix}} \def\ematrix{\end{matrix}} \def\bpmatrix{\begin{pmatrix}}\def\epmatrix{\end{pmatrix}}
\def\bcenter{\begin{center}} \def\ecenter{\end{center}}

\def\lowerheightfig#1#2#3{\(\raise-#1\hbox{\includegraphics[height=#2]{#3}}\)}
\def\lowerwidthfig#1#2#3{\(\raise-#1\hbox{\includegraphics[width=#2]{#3}}\)}

\usepackage{color}

 \publishversion

\begin{document}

\bcenter 
{\LARGE\textbf{The unbearable smallness of magnetostatic QCD corrections\\[25pt]}}
{\large Chris P. Korthals Altes}  \\[20pt]
National Institute for subatomic physics NIKHEF\\[5pt] Theory group\\[5pt]

Science Park 105\\[5pt] 
1098 XG, Amsterdam\\ [5pt]

The Netherlands  \\[10pt] 

and\\[10pt]

Centre de Physique ThŽorique\\[5pt]
Campus de Luminy, Case 907\\[5pt]
163 Avenue de Luminy\\[5pt]
13288 Marseille cedex 9, France\\[10pt]

\ecenter
\vfil
\begin{quotation} \noindent {\large\bf Abstract } \medskip \\
 One loop corrections to the magnetostatic QCD action are evaluated to  dimension six  in the magnetostatic fields.The result is remarkably simple

$${1\over 4}\int d\vec x ( F^a_{ij})^2-{g_M^2N\over{32\pi m_E^3}}\int d\vec x\Bigg({1\over{60}}(D_m^{ab}F^b_{mr})^2+{1\over{180}}f^{abc}F^a_{mn}F^b_{nr}F^c_{rm}\Bigg)(1+\OO(g_M^2)).$$
In all of the deconfined phase the correction is quite small.
 The term of $\OO(g^2_M)$ is the dimension  eight contribution
and is only presented in a  covariant derivative basis. We discuss their physical relevance for the pressure and the spatial Wilson loop. 
 \end{quotation}
\vfil 
\noindent Version: December 2017 \   \\[2pt]
{\footnotesize Last typeset: \today}
\eject
\setcounter{page}{2}

\section{Introduction}

The quark gluon plasma is experimentally analyzed at RHIC and LHC.  The underlying theory,
thermal QCD, has made great strides in elucidating its phase structure due to lattice simulations in four dimensions.  

In parallel a combination of dimensional reduction and lattice techniques has proven very useful. Dimensional reduction~\cite{dimreduction} is based on the idea to integrate out those degrees of freedom that are $\OO(T)$. In doing so one obtains the electrostatic QCD action (EQCD). This action is only valid if the electric screening
$m_E=\OO(g T)$  is much smaller than the scale $T$.  It lives in three dimensions since one is integrating out the $\OO(T)$ degrees of freedom. All Matsubara modes except the one that is constant in Euclidean time  are contributing {\it only} to the parameters in the EQCD action. So the fields in the new action have become time independent: $A_\m(\vec x), \m=0,1,2,3$.  $A_0$ is now a scalar field in three dimensions. The three spatial components of the vector potential appear on a different footing. They still constitute a three dimensional Yang-Mills theory.  Such a theory is not amenable to a perturbative treatment.  It has a magnetic screening scale $m_M=\OO(g^2 T)$ and Linde's argument~\cite{linde} show that to a given order depending on the dimension $d$ of the observable in question,  $g^{2d}$ or higher, an infinite number of diagrams contributes. 
Therefore the lattice simulations based on dimensional reduction are three dimensional, and take apart from EQCD only the MQCD action $\int d\vec x ~tr ~F_{ij}^2$ into account. 
 
Corrections to the MQCD action are obtained by integrating out the scale $m_E$.
This is still a meaningful expansion, although the corrections are not anymore a series in $g^2$ but in ${g_E^2\over{m_E}}=\OO(g)$.  It is these corrections we calculated to one loop order.
The lay-out of the paper is as follows. In section (\ref{sec:setup}) notation is fixed. In the next section (\ref{sec:sixeight})   we show the result for sixth and eighth order in the magnetostatic fields.  The last section discusses the results and their possible use.  For the necessary background the reader is invited to read the beautiful lecture notes by Laine and Vuorinen~\cite{mikko}.  The realization that corrections to the magnettostatic action were small is old~\cite{mikko},\cite{pierre}. What is new in this paper is the precise form of the correction.

\section{Effective action and dimensional reduction}\la{sec:setup}

Equilibrium quantities in QCD are described by Euclidean path integrals involving the QCD action.
Dimensional reduction consists of integrating out  the non-zero Matsubara modes $2\pi n T$. The modes  surviving are time independent. 

Below we give the result for the three dimensional electrostatic action $S_{E}$ obtained by the one loop approximation:
\be
S_{E}={1\over {g_E^2}}\left(\Tr F_{ij}^2+\Tr (\vec D(A)A_0)^2+ m_E^2 \Tr A_0^2+\l_E(\Tr A_0^2)^2 + \tilde{\l}_E\Tr(A_0^4)\right)
+\delta S_{E}.
\la{eqcd}
\ee
Vector potentials are denoted by e.g. $A_0=A_0^a  {\l^a\over 2}$, and field strengths $F_{ij}$ similarly. The $\l^a$ are the Gell-Mann matrices.
The covariant derivative is the adjoint one
\ba
\vec D(A)&=&\vec \partial+ T^b\vec A^b\non\\
T^b_{ac}&=&f_{abc}\non\\
~[\l^a,\l^b]&=&2if_{abc}\l_c,~tr\l_a\l_b=2\d_{ab}
\la{adjointcov}
\ea
$\Tr$ means trace over color and integration over space, $tr$ is trace over color only.
$\delta S_{EQCD}$ contains $\OO(A_0^6)$ terms. Taking these into account destroys the superrenormalizability of $S_{EQCD}$.

In this approach, when we integrate out hard ($\sim T$) degrees of freedom, we get  for the parameters in $EQCD$: 
\ba
g_E^{-2}&=&{1\over{g^2T}}\left(1+\#g^2+\#g^4+\#g^6+\cdots{....}    \right)\non\\
\l_E(\tilde\l_E)&=&T\left(\#g^4+\#g^6+\cdots{.....}\right)\non\\
m_E^2&=&\left(g^2NT^2/3+N_fg^2T^2/6\right)\left(1+\#g^2+\#g^4+\# g^6+\cdots{.....}\right).
\la{electricpara}
\ea

All indicated coefficients have been calculated (except for the last term in the coupling constant renormalization), for  reviews see~\cite{york15,york12,york12b}. 
The running coupling $g=g(T/\L_{\overline{MS}})$ is determined by four dimensional methods as expounded in~\cite{twogE}. Four dimensional lattice methods relating $\L_{\overline{MS}}$
to $T_c$ are found there as well.  As a result the  running coupling stays $\OO(1)$ at $T=T_c$.
  
 $N_f$ is the number of fermions involved. The fermion
fields are all heavy so their effect appears only in the thermal mass and the couplings. From Eq. (\ref{eqcd}) we read of the scale in EQCD to be the thermal mass $m_E=\OO(gT)$.
In contrast the scale in MQCD is $g_M^2=\OO(g^2T)$.  

\section{Integrating out soft modes. Magnetostatic QCD}

To obtain $S_M$  one integrates out degrees of freedom of the soft scale $m_E$.  This can be still be done in perturbation theory. It involves the field $A_0$, and in two loop or higher order
also the propagators and couplings in $F_{ij}^2$. The dimensionless expansion parameter is here $g_E^2/m_E=\OO(g)$.

Doing this leads to corrections to $Tr F_{ij}^2$:
\be
S_{M}&=&{1\over{g_M^2}}\Tr F_{ij}^2 +\D S_M.
\la{mqcd}
\ee

The magnetic coupling $g_M^2$ incorporates  now the well known one loop term in:

\be
{1\over{g_M^2}}={1\over{g_E^2}}\Bigg[1+{1\over 2}{g_E^2N\over{24\pi m_E}}
+{19\over 8}\Bigg({g_E^2N\over{24\pi m_E}}\Bigg)^2+....\Bigg].
\la{gmrenorm}
\ee

For completeness we have mentioned the two loop term calculated by Giovannangeli\cite{pierre2}. Doing so illustrates nicely the point that the expansion parameter is quite small:  
by choosing the expansion parameter  (somewhat arbitrarily) as ${1\over{24}}{g_E^2N\over{\pi m_E}}$ shows how the coefficient of the two loop term 
compares to that of the one loop term. Note that this expansion parameter is parametrically  $\OO(g)$, but is numerically small, even near $T_c$, as follows from the discussion after
Eq.(\ref{electricpara}) and the results for the gauge coupling  $g_E$ in ref.~\cite{twogE}. 


Fluctuations on the order of the thermal mass can be integrated out. The quadratic fluctuations in EQCD are lowest order, and give according to
Eq. (\ref{eqcd}) the determinant of the corresponding operator $-(\vec D(A))^2+m_E^2$.
 
\subsection{Lowest order terms in $S_{M}$}\la{sec:sixeight}
To obtain explicitly the corrections to the classical MQCD term $Tr F_{ij}^2$ one has to work out  the fluctuation determinant involving $A_0$ in the electrostatic action.
\be
\D S_M={1\over 2}\log\det(-\vec D^2+m_E^2)/(-\vec\partial^2+m_E^2)+\cdots{....}
\ee

We normalize by the determinant that gives the $\OO(m_E^3)$ contribution to the pressure.

The lowest order terms can be easily derived~\footnote{It is implicitly done by combining old work by Chapman~\cite{chapman} and  unpublished work by Giovannangeli~\cite{pierre}. }

The result is:
\ba
S_{M}&=&{1\over {g_E^2}}\Bigg[{1\over 4}\int d\vec x (F^a_{kl})^2-{g^2_E\over{192\pi m_E}}\Tr [D_kD_l]^2\non\\
&-&{g^2_E\over{32 \pi m_E^3}}\left({1\over {90}}\Tr [D_k D_l][ D_m D_n] [D_p  D_q]-{1\over{60}}\Tr [D_k[  D_k  D_l]]  [D_m [ D_m  D_l]]\right)\Bigg]\non\\
&=&{1\over{g_M^2}}\Bigg[{1\over 4}\int d\vec x ( F^a_{ij})^2\non\\&-&{g^2_MN\over{32\pi m_E}}\int d\vec x\Bigg({1\over{60}}({D_m^{ab}\over{}m_E}F^b_{mr})^2+{1\over{180}}f^{abc}{F^a_{mn}\over{m_E^2}}F^b_{nr}F^c_{rm}\Bigg)+....\Bigg]
\la{final}
\ea
The term quadratic in the field strength just renormalizes the coupling, as in Eq. (\ref{gmrenorm}).
So the relative order of $\D S_M$ with respect to $Tr F_{ij}^2$ is ${g^2_E\over{32 \pi m_E}}{[D_iD_j]]\over{m_E^2}}=\OO(g^3)$.   The next term with eight derivatives is $\OO([D_iD_j]/m_E^2)=\OO(g^2)$ smaller, and this is the general trend as shown in the next section below Eq. (\ref{expand}).

The relative correction to the classical magnetic field density is $\OO(g^3)$ (because the covariant derivative in the magnetic sector $D_i$ is$ \OO( g^2)$), and this correction contains only two dimension six invariants. 

\section{One loop correction with dimension six operators}

We start with the one loop result for the magnetic action :
\ba
S_{M}&=&{1\over {4g_E^2}}\int d\vec x ( F^a_{ij})^2+{1\over 2}\log\det (-D^2+m_E^2)/(-\partial^2+m^2)\non\\
&=&{1\over {4g_E^2}}\int d\vec x ( F^a_{ij})^2-{1\over 2}\int{dt\over t}\int {d\vec p\over {(2\pi)^3}}\Tr(\exp((D^2+2i\vec p.\vec D)t)-1)\non\\&\times&\exp(-(p^2+m^2)t)
\la{effaction}
\ea
\noindent and $D_i$ the adjoint covariant derivative:
\be
D_i=\partial_i{\bf 1}+T^cA_i^c, ~~T^c_{ab}=f^{acb}.
\la{covariantderivative}
\ee
The electric mass is $m_E^2=g^2NT^2/3+N_fg^2T^2/6$ and

\ba
[D_i,D_j]&=&F_{ij},~F_{ij}=T^aF^a_{ij}\non\\
tr T^aT^b&=&-N\d_{ab}\non \\
~[T^a,T^b]&=&f^{abc}T^c
\ea
and 
\be
tr T^aT^bT^c&=-{N\over 2} f_{abc}-{N\over 2}d^{abc}.
\la{adjointtrace}
\ee
\subsection{Expanding to order $t^3$}
There is only one explicit scale in the determinant, $m_E$. So taking it out gives us the factor $m_E^3$ familiar from the lowest order $S_{E}$ 
contribution to the pressure. 

It is useful to introduce a plane wave basis in the trace 
\be
\int{d\vec p\over {(2\pi})^3} \exp(-i\vec p.\vec y) D^2\exp{i\vec p.\vec x}=\int{d\vec p\over {(2\pi})^3} \exp(-i\vec p.(\vec y-\vec x))|(D^2+2i\vec p.\vec D-(\vec p)^2).
\la{planewavedet}
\ee

Then we let $y\rightarrow x$ and obtain for contribution of  the determinant:
\be
\D S_M=-{1\over 2}\int{dt\over t}\int {d\vec p\over {(2\pi)^3}}\Tr(\exp((D^2+2i\vec p.\vec D)t)-1)\exp(-(\vec p^2+m_E^2)t).
\la{expand}
\ee

The legitimacy of the expansion of the determinant is based on the fact that in the magnetic sector the ratio of magnetic over electric scales appears in the covariant derivatives:
\be
D/m_E=\OO(g).
\ee
Rotational invariance allows only terms with an even number of derivatives $D$.  For a given degree (starting with $D^6$) the sum can be rewritten as a sum of terms, each consisting of products of multiple commutators of $D$'s, like $[D_iD_j]=F_{ij}$ and so on.  

The first step is to obtain the terms written as strings of covariant derivatives. This is  simple and straightforward  using the formulas in the Appendix.

The trace in Eq. (\ref{expand})  allows to perform a cyclic permutation on any given string of covariant derivatives.  This limits all fourth order terms to
be either $\Tr D^4$ or $\Tr (D_iD_j)^2$.  There are five sixth order terms, shown below in Eq. (\ref{5terms}). 

Explicitly  for terms of up and including order $t^3$ we find the following. 

 The terms of order $t$ in the expansion cancel.

 Next are the two $t^2$ terms. The first one is obtained by expanding the exponential 
 \be
 \int{d\vec p\over{(2\pi)^3}}\exp((D^2-2i\vec p.\vec D)t-p^2t)
 \ee
 to $\OO(t^2)$, the second to $\OO(t^4)$  doing the $\vec p$ integration using Eq. (\ref{akcoeff}):
 \ba
 -{1\over{12}}t^2\Tr D^4I(t)
 +{1\over{12}}t^2\Tr (D_kD_l)^2I(t)
 \ea
 and the sum is 
 \be
{1\over {24}} t^2\Tr [D_k,D_l]^2I(t).
\la{2terms}
 \ee
 So we have recovered a term proportional to the classical action. This term corresponds of course to the renormalization of the magnetic coupling, the one loop term in Eq. (\ref{gmrenorm}).
 The five $t^3$ terms are, following the same lines
 \ba
{1\over 180}t^3I(t)&\times&\Bigg(-2\Tr D^6-3\Tr D^2D_kD^2D_k+9\Tr (D_kD_l)^2D^2\non\\
                                                 &-&\Tr D_kD_lD_mD_kD_lD_m-3\Tr D_kD_lD_mD_lD_kD_m\Bigg)
\la{5terms}
 \ea

If the covariant derivatives would commute the first three terms cancel with the last three terms in Eq.(\ref{5terms}).
Therefore the sum can be rearranged into  terms with at least one commutator. As we expect the outcome to depend only on the gauge field configuration any term can only depend on (multiple) commutators. In this case it turns out that one can rewrite the contribution in parenthesis  as the sum
of only two terms:  one being the square of the operator of the equation of motion, $D_kF_{kl}$  (see Appendix). The second term is  cubic in the magnetic field strengths:
\ba
&&-2\Tr D^6-3\Tr D^2D_kD^2D_k+9\Tr (D_kD_l)^2D^2 -\Tr D_kD_lD_mD_kD_lD_m\non\\
&&-3\Tr D_kD_lD_mD_lD_kD_m\non\\
&=&\Tr[D_kD_l][D_lD_m][D_mD_k]-{3\over 2} \Tr ([D_k[D_kD_l]])^2\non\\
&=&- {N\over 2}f^{abc}\vec B^a.\vec B^b\times\vec B^c+{3N\over 2}(\vec D^{ab}\times \vec B^b)^2,
\la{5term1}
\ea
\noindent using Eq. (\ref{adjointtrace}) whilst  the outer product
\be
(\vec x\times\vec y)_k\equiv \epsilon_{klm}x_ly_l,~\epsilon_{123}=1.
\ee

This leads to the result for $\D S_M$ as mentioned in Eq. (\ref{final}).

\subsection{Dimension eight  corrections}

A straightforward calculation of the eighth order  polynomial using (\ref{akcoeff}) leads to an expression with eighteen  independent invariants, instead of five as for the sixth order. These invariants 
are written in a short hand form. This is best understood by giving an example .
Take a term:
\be
tr D_jD_jD_kD_lD_mD_lD_kD_m=(0312).
\la{numbernotation}
\ea
We start with the first index j. It must contract with another index j. This index is separated from the first one by no other $D_x$, and so we start with 0. Next index is k and it contracts  with another index k, that is separated by three $D_x$, so the next number is 3. The next index l is separated by one derivative $D_m$, and the last index m by 2
derivatives. This gives the result   as in Eq. (\ref{numbernotation}).
Because of the cyclical trace it can also be written as (3120) or  (2530), but we take the convention that the smallest number comes first. Because of the cyclic property any number is modulo 6, hence (2530)=(0312). 

One finds by inspection that there are 18 independent terms as in (\ref{numbernotation}). All other combinations reduce to those 18 terms, through cyclicity.

Now we can write the expansion of the determinant to eighth order.
\ba
\log\det (D^2+m_E^2)|_{8th}&=&\int_0^\infty{dt\over t}{I(t)\over{7!}}\Bigg[-10 (0000)+ 72 (0011) -68 (0020) -32 (0121)\non\\&+& 72 (0130)  -40 (0222)  -40 (0231) -44 (0312)+  32 (0330) +36 (0420)  -40 (0411)\non\\
&+&8 (1111)+16 (1221) +16 (1322)+8 (1331)+8 (2332) +4 (2422) +2 (3333)\Bigg]
\la{eightorderstring}
\ea

As for the sixth order result the coefficients in the first line and those in the second line do add up to zero. So also here the result consists of terms with at least one commutator factor.   We have  determined the minimal basis of mulitiple commutator terms like we did for the sixth order case. Here again "minimal" means that any other multiple 
commutator term can be written through cyclicity of the trace and using the Jacobi identity as a linear combination out of the minimal set. The result is given in the Appendix.

%
\section{Use in the EOS}

Let us see what corrections the free energy picks up from the six order terms in Eq. (\ref{final}).
We treat the correction as a perturbation,   so the theory stays superrenormalizable.
So without the correction the free energy is of the form
\be
f_M V/T=-\log \int D\vec A\exp(-{1\over {g_E^2}}\int d\vec x S_M)=c(g_E^2T)^3 V/T.
\la{eoslowest}
\ee 

Including $\D S_M$ we get
\be
f_M V/T=-\log \int D\vec A\exp(-{1\over {g_E^2}}S_M)(1-\D S_M)=-\log \int D\vec A\exp(-{1\over {g_E^2}}S_M)+\langle\D S_M\rangle.
\ee

The last term equals because of translation invariance
\be
\langle\D S_M\rangle=V\langle\D S_M(0)\rangle.
\ee

Finally the expectation value of $\D S_M(0)$ has to be measured on the lattice. Because it is of sixth order in the covariant derivative we expect for the average
\be
\langle tr B_iB_jB_k\epsilon_{ijk}\rangle=c_1g_M^{12}, c_1~\mbox{a dimensionless lattice constant}
\ee
and similar for $tr(D_iF_{ij})^2$ with a coefficient $c_2$ .  The presence of the factor ${1\over{m_E^3}}$ in $\D S_M$ renders the effect $\OO(g^9)$ so as already mentioned $\OO(g^3)$
smaller than $f_M$.

\section{Use in the spatial string tension}

In four dimensional QCD at a finite temperature we can define a Wilson loop in a given time slice. Such a spatial loop, $tr \exp(i\oint\vec A.d\vec l)$, will in the Abelian case 
through Stokes law catch the magnetic flux in the plasma (more precisely, fluctuations thereof).  However in that case there is no reason to expect any magnetic flux at all, and the loop will obey the  perimeter
law, reflecting the radiative corrections to the $\vec A$ potential.


 The  spatial string tension is given by a rectangular spatial Wilson loop, say in the x-y plane, of size $L\times L$:
 \be
 W(L)=\langle  tr{\cal P}\exp( i\oint_L\vec A.\d\vec l) \rangle
 \la{wline}
 \ee
For definiteness we take as gauge group $SU(3)$, and $N_f\neq 0$. The average is the usual thermal path integral average.
The spatial tension is then obtained from the loop average by
\be
\sigma_s(T)=-\lim_{L\rightarrow\infty}\log W(L).
\la{spacetension}
\ee
This tension has been  measured on the lattice in 4d with and without fermions and in 3d pure gauge fields~\cite{boyd}. 

At very high temperature the tension will pick up only large distance contributions from the magnetic action $S_M$ after integrating out scales $T$ and $gT$.

The path integral average is now 3 dimensional and the large distance contributions are described by a non-perturbative coefficient $c_s $ multiplying
$g_M^4$ for dimensional reasons,
\be
\sqrt{\sigma_s(T)}=c_s g_M(T)^2.
\la{tensiongm}
\ee

If we neglect $\D S_M$ we have for $c_s$ from 3d lattice simulations~\cite{boyd}
\be
c_s(0)=0.553(1),~ N=3.
\ee

Taking  $\D S_M$  into account we expand the exponential $\exp(-S_M-\D S_M)$ in the path integral for the Wilson loop average
\ba
&&{\int D\vec A W(L)\exp(-S_M-\D S_M)\over{\int D\vec A \exp(-S_M-\D S_M)}}={\int D\vec A W(L)\exp(-S_M)(1-\D S_M)\over{\int D\vec A \exp(-S_M)(1-\D S_M)}}\non\\
&=&{\int D\vec A W(L)\exp(-S_M)(1+\langle\D S_M\rangle)\over{\int D\vec A \exp(-S_M)}}-{\int D\vec A W(L)\D S_M(1+\langle\D S_M\rangle)\exp(-S_M)\over{\int D\vec A\exp(-S_M)}}\non\\
&=&\langle W(L)\rangle+\langle W(L)\langle\D S_M\rangle\rangle- \langle W(L) \D S_M\rangle)+ \OO(\D {S_M}^2)\non\\
&=&\langle W(L)\rangle\left(1- {\langle W(L) \D S_M\rangle_c\over{\langle W(L)\rangle}}\right).
\ea
This involves the truncated correlation of the loop with the correction to the action.
From this we find easily, using Eq. (\ref{spacetension}) that 

\be
\s_s=\s_s(0)+\D\s_s,
\ee
\noindent with
 
\be
\D\s=-\lim_{L\rightarrow\infty}{1\over{L^2}}{\langle W(L)\D S_M)\rangle_c\over{\langle W(L)\rangle}},
\la{correlationdeltasm}
\ee

\noindent and can only be obtained from lattice simulations~\footnote{On the lattice the  strong coupling expansion produces indeed an area law for the right hand side of Eq. (\ref{correlationdeltasm}). For the operator $tr F^3$ the strong coupling series starts with a term $\b_L^3$ in the lattice coupling.}.  
We expect $\OO(g^3)$ corrections to $c_s(0)$, with again a very small coefficient.
~\\
~\\
~\\
\section{Conclusions}

We have  derived the one loop corrections to the magnetostatic action by integrating out fluctuations on the screening scale $m_E$. They are the product of the renormalization of the electrostatic coupling to the magnetostatic coupling $g_E^2N/(32\pi m_E)$ and a sum of dimension six, eight,....operators in units of $m_E^2$, $m_E^4$,.....  The first term in the sum is $\OO(g_E^2)$, the second $\OO(g_E^4)$ and so on.
The first factor is quite small in the plasma phase\cite{pierre3}. Pressure and spatial string tension are only very little affected. For the renormalization of $g_M$  in terms of $g_E$ this has been analyzed and corroborated in ref.\cite{pierre3}.
It is perhaps worthwhile to note that our  Skyrme-like terms of dimension six  are not generating a classical stable solution. The reason is that the dominant term $\Tr F_{ij}^2$ has qualitatively the same scaling properties as our correction terms, i.e. they scale all with a positive power. This fact excludes a non-trivial minimum.

\section{Acknowledgements}
I am indebted to the theory groups of the CPT and NIKHEF for hospitality. Mikko Laine and Pierre Giovannangeli have given sound advice.

\section{Appendix: Independent six and eighth order operators}\ref{sec:eighthorder}

Below we derive the formulas needed for the calculation of  the corrections.
Due to the presence of the term $\vec p.\vec D$ in  (\ref{expand}), we need some notation:
\ba
\langle p_ip_j\cdots\rangle&\equiv& \int_{-\infty}^\infty{d\vec p\over{(2\pi)^3}}p_ip_j\cdots\exp(-\vec p^2t)\\
\langle 1\rangle&=&\Bigg({1\over{4\pi t}}\Bigg)^{{3\over2}} \equiv I(t)
\ea
and

\ba
a_2 \d_{ij}&\equiv& t(2i)^2<p_ip_j>, ~a_2=-2 I(t)\non\\
 a_4(\d_{ij}\d_{kl}+\mbox{2 perm})&\equiv&t^2(2i)^4<p_ip_jp_kp_l>,  ~a_4=4 I(t)\non\\
 a_6(\d_{ij}\d_{kl}\d_{mn}+\mbox{14 perm})&\equiv&t^3(2i)^6<p_ip_jp_kp_lp_mp_n>, ~a_6=- 8 I(t)\non\\
  a_8(\d_{ij}\d_{kl}\d_{mn}\d_{rs}+\mbox{104 perm})&\equiv&t^4(2i)^8<p_ip_jp_kp_lp_mp_np_rp_s>, ~a_8=16 I(t)
\la{akcoeff}
\ea
\ba
e_{i_1, \cdots, i_{2n}}^2&=&2n-1!! 2n+1!!\non\\
(2i)^{2n}e_{i_1,\cdots,i_{2n}} \langle p_{i_1}\cdots p_{i_2n}\rangle&=&(-2)^n2n-1!!2n+1!! I(t)\non\\
\langle {\vec p}^{2n}\rangle&=&(-{d\over{dt}})^nI(t)={(2n+1)!!\over{2^n}}{I(t)\over{t^n}}\non\\
\mbox{hence:} ~a_{2n}(t)=(-2)^nI(t).
\ea

The result of the effective action is written in strings of covariant derivatives as Eq.(\ref{5terms}) and (\ref{eightorderstring}). But it is ultimately in terms of field strengths and their covariant derivatives.
Therefore the result can be written in terms of factors, each of which is a multiple commutator in terms of covariant derivatives. The overall color trace 
necessitates at least two of such factors.
The contraction of indices should give a scalar, we have cyclic permutation inside the trace, and the Jacobi identity:
\ba
~[D_i[D_jD_k]]&=&[[D_iD_j]D_k]+[D_j[D_iD_k]]\non\\
~[D_i[D_j[D_kD_l]]]&=&[[D_iD_j][D_kD_l]]+[D_j[[D_iD_k]D_l]]+ [D_j[D_k[D_iD_l]]].
\ea

To order four we only one possibility:
\begin{itemize}
\item $Tr [D_iD_j]^2$.
\end{itemize}
 
 To order six we have three structures possible:
 \begin{itemize}
 \item $C_1=Tr[D_iD_j][D_i[D_k[D_jD_k]]]$ (1)
 \item $C_2=Tr[D_k[D_kD_i]]^2 $ (1)
 \item $C_3=Tr[D_i[D_jD_k]]][D_i[D_jD_k]] $ (0) 
 \item $C_4=Tr[D_iD_j][D_kD_i][D_jD_k][$ (1).
 \end{itemize}
 $C_3$ can be trivially rewritten as a $C_1$ or $C_2$ type.
Only the second and the fourth contribute to the determinant.

To order eight  the following structures are possible-writing for short  $[D_iD_j]=F_{ij}$:
\begin{itemize}
\item $O_1= Tr F_{ij}F_{jk}F_{kl}F_{li}$ and $Tr F_{ij}F_{ji}F_{kl}F_{lk}$ (2)
\item $O_2=Tr F_{ij}F_{kl}[D_i[D_jF_{kl}]]$ (1)
\item $O_3=Tr F_{ij}F_{jl}[D_i[D_kF_{kl}]]$  (2)
\item $O_4=Tr F_{ij} [D_lF_{li}] [D_mF_{mj}]$ (1)
\item $O_5=Tr F_{ij} [D_lF_{lm} [D_mF_{ij}]$  (2)
\item $O_6= Tr [D_i[D_jF_{kl}]] [D_i[D_jF_{kl}]]$  (6) 
\item $O_7= Tr [D_i[D_kF_{kj}]] [D_i[D_lF_{lj}]]$  (2)
\item $O_8= Tr [D_i[D_kF_{kj}]] [D_l[D_lF_{ij}]]$  (1)
\end{itemize}
In $O_2$ and $O_6$ all permutations of the 4 indices in say the righthand factor are understood.

 We see that  transposition of the indices $i$ and $j$ in $O_6$ leads to an operator of type $O_2$ and $O_4$.  And obviously transposition of $k$ and $l$
 just flips the sign. So out of the 24 permutations that constitute $O_6$ only 6 are independent.  
$O_1$ has only two inequivalent contractions. $O_4$ is unique.


\begin{thebibliography}{99}

\bibitem{dimreduction}T.  Appelquist  and  R.D.  Pisarski,
High-temperature  Yang-Mills  theories  and three-
dimensional Quantum Chromodynamics,
Phys. Rev. D 23 (1981) 2305.
\bibitem{linde}A.D. Linde,
Infrared problem in thermodynamics of the Yang-Mills gas,
Phys. Lett. B 96
(1980) 289.
\bibitem{mikko}
Basics of Thermal Field Theory,
Mikko Laine, Aleksi Vuorinen.
 Lect.Notes Phys. 925 (2016) 
\bibitem{chapman}S. Chapman, Phys.Rev. D50 (1994) 5308-5313.
\bibitem{pierre} Pierre Giovannangeli, Seminar notes Trento September 2005.

 \bibitem{york15} 
Ioan Ghisoiu, Jan Moller, York Schroder , JHEP 1511 (2015) 121,: arXiv:1509.08727 [hep-ph] 
\bibitem{york12} 	
J. Moller, Y. Schroder, JHEP 1208 (2012) 025
, arXiv:1207.1309 [hep-ph] | PDF

\bibitem{york12b}  	
Jan Moller, York Schroder, Prog.Part.Nucl.Phys. 67 (2012) 168-172

\bibitem{boyd} 	
G. Boyd, J. Engels, F. Karsch, E. Laermann, C. Legeland, M. Lutgemeier, B. Petersson,  Nucl.Phys. B469 (1996) 419-444
, hep-lat/9602007;
M. Cheng (Columbia U.) et al.,  Phys.Rev. D78 (2008) 034506
; PhysRevD.78.034506; arXiv:0806.3264 [hep-lat] .
\bibitem{pierre2} P. Giovannangeli, Phys.Lett. B585 (2004) 144-148,
hep-ph/0312307.
\bibitem{pierre3}Two loop renormalization of the magnetic coupling and non-perturbative sector in hot QCD
P. Giovannangeli, Nucl.Phys. B738 (2006), 23.
\
\bibitem{laine1}M. Laine and Y. Schroder,
Two-loop QCD gauge coupling at high temperatures,
JHEP 03
(2005) 067 [hep-ph/0503061].
\bibitem{twogE} M. Laine and Y. Schroder, JHEP
0503
(2005) 067 [arXiv:hep-ph/0503061].

\end{thebibliography}
\end{document}